\documentclass[12pt]{article}
\usepackage{graphicx}
\usepackage{amsmath}

\numberwithin{equation}{section}

\begin{document}

\title{A Random Matrix Models of Relaxation}
\author{J. L.\ Lebowitz \\
Department of Mathematics, Rutgers University, USA \thanks{%
Also Department of Physics of Rutgers University} \and L. Pastur \\
Department of Mathematics, University Paris 7, France \thanks{%
Also Mathematical Division, Institute for Low Temperature Physics, Kharkov,
Ukraine} }
\date{}
\maketitle

\begin{abstract}
\noindent We consider a two level system, $\mathcal{S}_{2}$, coupled to a
general $n$ level system, $\mathcal{S}_{n}$, via a random matrix. We derive
an integral representation for the mean reduced density matrix ${\rho} (t)$
of $\mathcal{S}_{2}$ in the limit $n\rightarrow \infty $, and we identify a
model of $\mathcal{S}_{n}$ which possesses some of the properties expected
for macroscopic thermal reservoirs. In particular, it yields the Gibbs form
for ${\rho} (\infty )$. We consider also an analog of the van Hove limit and
obtain a master equation (Markov dynamics) for the evolution of $\rho (t)$
on an appropriate time scale.

\bigskip \noindent
PASC numbers: \;  05.30.Ch, 05.60.Gg, 05.70.Ln 

\end{abstract}

\section{Introduction}

\label{s:intr}

\noindent The time evolution of a "small" quantum system interacting with a
\textquotedblright big\textquotedblright\ thermal reservoir is a much
studied problem in various contexts, see review works \cite%
{Ku-Co,Ha,Da,Li-Co,Sp,Ja-Pi:97,Ba-Co}. Two natural questions for such models
are the large time limit of the state of the small system and the nature of
the approach to this limit. It is generally expected and rigorously proven
for certain types of reservoirs and couplings that, starting from an
arbitrary initial state of the system and an equilibrium Gibbs state at
temperature $T$ of the reservoir, the final, $t\rightarrow \infty $, state
of the system is given by a projection of the joint system plus reservoir
Gibbs measure at the temperature $T$\cite{Ba-Co,Ja-Pi:97}.

In this paper we discuss the above questions in the frameworks of random
matrix theory. Namely, we consider a small system $\mathcal{S}_{2}$, having
only two levels (like in the archetype spin-boson model \cite{Sp,Ba-Co,
Ja-Pi:97}), coupled to a system with $n$ levels (denoted by $\mathcal{S}_{n}$%
) via an $n\times n$ random matrix. Similar models have been discussed in
papers \cite{Bu,Me-Co,Pe,Lu-We}. Here we will investigate in a more explicit
way what features of the $n$-level system will lead, in the limit $%
n\rightarrow \infty $, to the reduced density matrix $\rho (t)$ of the small
system having a Gibbs form for large times and whether $\rho (t)$ is
described by a Markov evolution. Our analysis will be based entirely on the
fact that the interaction is given by a typical realization of a random
matrix. This models in some way a strongly interacting reservoir.%

The paper is organized as follows. In Section 2 we describe the model and
define the reduced density matrix for the small, two level, system. Since it
is random, its full statistical description is of interest. In this paper we
confine ourselves to the analysis of its mean which we argue is also its
typical behavior in certain cases. 
In Section 3 we consider the equilibrium properties of the
combined system. We find the reduced density matrices of
$\mathcal{S}_{2}$ corresponding to the microcanonical and
canonical description of the composite system and give conditions
for their equivalence.

Section 4 deals with time dependent (non-equilibrium) properties of the
model. We
find that the $t\rightarrow \infty $ limit of ${\rho}(t)$ depends in general
on the initial density matrix. However, for the reservoir introduced in
Section 3 designed to guarantee the equivalence of the microcanonical and
canonical description in equilibrium, the $t\rightarrow \infty $ limit of
the diagonal entries of the mean reduced density matrix is independent of
the initial density matrix and has the Gibbs form. In Section 5 we treat the
van Hove asymptotic regime (small coupling, long times). Here both the
equilibrium and the time dependent form of the mean reduced density matrix
can be expressed directly via the density of states of the reservoir. We
find that for the special form of the density of states of $\mathcal{S}_n$
of Section 3 the 
$t\rightarrow \infty $ limit of ${\rho} (t)$ is the Gibbs distribution, and
its time evolution is Markovian. However, in this case the obtained time
decay rates prove to be vanishing. 
In Section 6 we consider a more general model, where one can obtain quite
reasonable decay rates. The model is similar to that, studied in \cite{Lu-We}%
. In Section 7 we continue studies of the time evolution $\rho(t)$
discussing its dependence on the initial density matrix and on the
parameters. Several formulas, used in the main body of the paper, are
derived in Appendices.


\section{Description of Model}

\label{s:mod}

\noindent Let $h_{n}$ be a real symmetric $n\times n$ matrix with
eigenvalues $E_{j},\;j=1,...,n$. We
characterize the spectrum of $h_{n}$ by its density of states
\begin{equation}
\nu _{0}^{(n)}(E)=n^{-1}\sum_{j=1}^{n}\delta (E-E_{j}).  \label{non}
\end{equation}%
We assume that $\nu _{0}^{(n)}$ converges as $n\rightarrow \infty $ to a
limiting density $\nu _{0}$, i.e.\ that for any continuous and rapidly
decaying function $\varphi $ we have:
\begin{equation}
\lim_{n\rightarrow \infty }\int_{-\infty }^{\infty }\varphi (E)\nu
_{0}^{(n)}(E)dE=\int_{-\infty }^{\infty }\varphi (E)\nu _{0}(E)dE,\quad
\;\int_{-\infty }^{\infty }\nu _{0}(E)dE=1.  \label{no}
\end{equation}%
Let $w_{n}$ be a real symmetric $n\times n$ random matrix, whose probability
density is
\begin{equation}
Q_{n}^{-1}\exp \left\{ -\frac{1}{4}\mathrm{Tr\;}w^{2}\right\} ,  \label{GUE}
\end{equation}%
where $Q_{n}$ is the normalization constant. In other words, the entries $%
w_{jk},\;1\leq j\leq k\leq n$ of the matrix $w_{n}$ are independent Gaussian
random variables, with
\begin{equation}
\left\langle w_{jk}\right\rangle =0,\quad \;\left\langle
w_{jk}^{2}\right\rangle =(1+\delta _{jk}),  \label{GOE}
\end{equation}%
where the symbol $\left\langle ...\right\rangle $ denotes averaging with
respect to the distribution (\ref{GUE}). This probability distribution is
known as the Gaussian Orthogonal Ensemble (GOE) \cite{Me}.

We define the Hamiltonian of our \textit{composite system }$\mathcal{S}%
_{2,n} $ as a random $2n\times 2n$ matrix of the form
\begin{equation}
H_{n}=s\sigma ^{z}\otimes \mathbf{1}_{n}+\mathbf{1}_{2}\otimes h_{n}+v\sigma
^{x}\otimes w_{n}/n^{1/2},  \label{Ham}
\end{equation}%
where $\mathbf{1}_{l}$ ($l=2,n$) is the $l\times l$ unit matrix, $\sigma
^{z},$ and $\sigma ^{x}$ are the Pauli matrices
\begin{equation*}
\sigma ^{x}=\left(
\begin{array}{ll}
0 & 1 \\
1 & 0%
\end{array}%
\right) ,\;\sigma ^{z}=\left(
\begin{array}{ll}
1 & 0 \\
0 & -1%
\end{array}%
\right) .
\end{equation*}%
The symbol $\otimes $ denotes the tensor product, and $s$ and $v$ are
positive constants.

The first term in (\ref{Ham}) is the Hamiltonian of $\mathcal{S}_{2}$, the
second term is the Hamiltonian of $\mathcal{S}_{n}$, and the third term is
an interaction between them. Thus $s$ determines the energy scale of the
isolated small system, and $v$ plays the role of the coupling constant
between the two level system $\mathcal{S}_{2}$ and the $n$ level system $%
\mathcal{S}_{n}$. 
Note also the $n^{-1/2}$ scaling of the interaction.

We emphasize that letting
$n\rightarrow \infty $ does not imply that $\mathcal{S}_{n}$ will behave
like a macroscopic system. $\mathcal{S}_{n}$ could perfectly well be just a
single particle moving on a periodic chain of $n$ sites with spacing $\delta
$, for which, in the limit $n\rightarrow \infty $,
\begin{equation}
\nu _{0}(E)=\left( \pi \sqrt{E(4\delta ^{-2}-E)}\right) ^{-1},\;0<E<4\delta
^{-2}.  \label{DOSE}
\end{equation}%
In fact the normalization assumption (\ref{no}) implies that we are not
dealing here with what one usually thinks of as a thermal reservoir.

We are going to study the time evolution of the \textit{mean} \textit{%
reduced density matrix} of $\mathcal{S}_{2}$, assuming that at $t=0$ the
density matrix of the composite system $\mathcal{S}_{2,n}$ is of the form
\begin{equation}
\mu _{m}^{(n)}(0)=\rho (0)\otimes P_{k},  \label{ini}
\end{equation}%
where $\rho (0)$ is a $2\times 2$ positive definite matrix of unit trace and
$P_{k}$ is the projection on the state of energy $E_{k}$ of the reservoir.
Let
\begin{equation}
\mu ^{(n)}(t)=\{\mu _{\alpha j,\beta k}^{(n)}(t),\;\alpha ,\beta =\pm
,\;j,k=1,...,n\}  \label{dmcom}
\end{equation}%
be the density matrix of the composite system $\mathcal{S}_{2,n}$ at time $t$%
, corresponding to the initial density matrix $\mu _{m}^{(n)}(0)$ of (\ref%
{ini}). Then the mean reduced density matrix of the small system is defined
as
\begin{equation}
{\rho }_{\alpha ,\delta }^{(n)}(E_{k},t)=\sum_{j=1}^{n}\left\langle \mu
_{\alpha j,\delta j}^{(n)}(t)\right\rangle =\sum_{j=1}^{n}\left\langle
\left( e^{-itH_{n}}\mu _{m}^{(n)}(0)e^{itH_{n}}\right) _{\alpha j,\delta
j}\right\rangle .  \label{rored}
\end{equation}%
We have evidently for all $k=1,...,n$ and $t\geq 0$ :
\begin{equation}
\sum_{\alpha =\pm }{\rho }_{\alpha ,\alpha }^{(n)}(E_{k},t)=1.  \label{rosr}
\end{equation}%
The mean reduced density matrix is determined by the initial density matrix $%
\rho (0)$ of $\mathcal{S}_{2}$ via the linear relation
\begin{equation}
{\rho }_{\alpha ,\delta }^{(n)}(E_{k},t)=\sum_{\beta ,\gamma =\pm }T_{\alpha
\beta \gamma \delta }^{(n)}(E_{k},t)\rho _{\beta ,\gamma }(0),  \label{ron}
\end{equation}%
where
\begin{equation}
T_{\alpha \beta \gamma \delta }^{(n)}(E_{k},t)=\sum_{j=1}^{n}\left\langle
\left( e^{-itH_{n}}\right) _{\gamma k,\delta j}\left( e^{itH_{n}}\right)
_{\alpha j,\beta k}\right\rangle ,  \label{per}
\end{equation}
is the \textquotedblright transfer\textquotedblright\ matrix. It
can be viewed as an analogue of the influence functional,
introduced by Feynman and Vernon in a similar context \cite{Fe}.

A particular case of the mean reduced density matrix is the mean transition
probability ${p}_{\alpha ,\delta }^{(n)}(E,t)$ for $\mathcal{S}_{2}$ to be
found in the state $\left\vert \alpha \right\rangle $ at time $t$ provided
that at $t=0$ it was in the state $\left\vert \beta \right\rangle $ and the
initial state of $\mathcal{S}_{n}$ was the pure state of energy $E_{k}$ for
some $k$. It is given by
\begin{equation}
{p}_{\alpha ,\delta }^{(n)}(E_{k},t)=T_{\alpha \delta \delta \alpha
}^{(n)}(E_{k},t)=\sum_{j=1}^{n}\left\langle \left\vert \left(
e^{-itH_{n}}\right) _{\alpha j,\delta k}\right\vert ^{2}\right\rangle ,
\label{ver}
\end{equation}%
with
\begin{equation}
\sum_{\alpha =\pm }p_{\alpha ,\delta }^{(n)}(E_{k},t)=1,\;\delta =\pm .
\label{psr}
\end{equation}%
The diagonal entries $p_{\alpha ,\alpha }^{(n)}(E_{k},t),\;\alpha =\pm $ are
known as \textit{survival (or return) probabilities}.


We can evidently replace $P_{k}$ by any superposition $\sum_{k}c_{k}P_{k}$
of pure states of ${}\mathcal{S}_{n}$, e.g.\ by that, corresponding to the
canonical density matrix $e^{-\beta h_{n}}\left/ Z_{n}\right. $, where $Z_{n}
$ is the partition function of ${}\mathcal{S}_{n}$ and $\beta $ is its
inverse temperature. The initial density matrix for the composite system
will now be (cf (\ref{ini}))
\begin{equation}
\mu _{c}^{(n)}(0)=\rho (0)\otimes e^{-\beta h_{n}}\left/ Z_{n}\right. ,
\label{cini}
\end{equation}%
corresponding to the choice
\begin{equation}
c_{k}=e^{-\beta E_{k}}\left/ n\int_{-\infty }^{\infty }e^{-\beta E}\nu
_{0}^{(n)}(E)dE\right. ,  \label{cini1}
\end{equation}%
where $\nu _{0}^{(n)}$ is defined in (\ref{non}).

It should be mentioned, that our results remain valid also in the case,
where the matrix $h_{n}$ is non-diagonal, random, independent of $w_{n}$,
and possesses property (\ref{no}) with probability 1. However, the
randomness of $h_{n}$ plays practically no role. Besides, our results are
not limited to the special Gaussian distribution (\ref{GUE}) - (\ref{GOE})
of the interaction, but hold also for any real symmetric random matrix $%
w_{n} $ in (\ref{Ham}), whose entries $w_{jk},\;1\leq j\leq k\leq n$ are
independent, verify (\ref{GOE}) and have fourth moments bounded uniformly in
$n$. The techniques in this case are, however, more involved.


\section{Equilibrium States}

\label{s:equi}

\noindent \medskip\textit{3.1. General formulas}.

\noindent We begin by considering the equilibrium (time independent)
microcanonical density matrix of the composite system ${}\mathcal{S}_{2,n}$:
\begin{equation}
\Omega (\lambda )=\delta (\lambda -H_{n})\left/ \mathrm{Tr}\;\delta (\lambda
-H_{n})\right. ,  \label{med}
\end{equation}%
Following a standard prescription of statistical mechanics, we will replace
the Dirac delta-function in (\ref{med}) by the function $(2\varepsilon
)^{-1}\chi _{\varepsilon }$, where $\chi _{\varepsilon }$ is the indicator
of the interval $(-\varepsilon ,\varepsilon )$, and $\varepsilon \ll \lambda
$. Then the reduced microcanonical density matrix of $\mathcal{S}_{2}$,
i.e., the microcanonical density matrix of $\mathcal{S}_{2,n}$, traced with
respect to the states of ${}\mathcal{S}_{n}$, is the $2 \times 2$ matrix of
the form
\begin{equation}
\omega ^{(n)}(\lambda )=\frac{\overline{\nu }^{(n)}(\lambda )}{\sum_{\delta
=\pm }\overline{\nu }_{\delta ,\delta }^{(n)}(\lambda )},  \label{mer}
\end{equation}%
where
\begin{equation}
\overline{\nu }_{\alpha \gamma }^{(n)}(\lambda )=(2\varepsilon
n)^{-1}\sum_{j=1}^{n}\chi _{\varepsilon }(\lambda -H_{n})_{\alpha j,\gamma
j}.  \label{mer1}
\end{equation}%
The corresponding canonical distribution of the composite system is
\begin{equation}
e^{-\beta H_{n}}\left/ \mathrm{Tr}\;e^{-\beta H_{n}}\right. ,  \label{ced}
\end{equation}%
and the reduced distribution of the small system is
\begin{equation}
\frac{\int_{-\infty }^{\infty }e^{-\beta \lambda }\nu ^{(n)}(\lambda
)d\lambda }{\sum_{\delta =\pm }\int_{-\infty }^{\infty }e^{-\beta \lambda
}\nu _{\delta ,\delta }^{(n)}(\lambda )d\lambda },  \label{cer}
\end{equation}%
where (cf (\ref{mer1}))
\begin{equation}
\nu _{\alpha \gamma }^{(n)}(\lambda )=n^{-1}\sum_{j=1}^{n}\delta (\lambda
-H_{n})_{\alpha j,\gamma j}.  \label{nuen}
\end{equation}%
By using standard techniques of random matrix theory (see e.g. \cite{P1,P2}%
), it can be shown that with probability 1 with respect to the distribution (%
\ref{GUE}) of the interaction matrix the $n\rightarrow \infty $ limit of (%
\ref{mer1}) is
\begin{equation}
\overline{\nu }_{\alpha \gamma }=\delta _{\alpha ,\gamma }\overline{\nu }%
_{\alpha },\;\;\overline{\nu }_{\alpha }(\lambda )=(2\varepsilon
)^{-1}\int_{\lambda -\varepsilon }^{\lambda +\varepsilon }\nu _{\alpha }(\mu
)d\mu  \label{nuab}
\end{equation}%
with
\begin{equation}
\nu _{\alpha }(\lambda )=\pi ^{-1}\Im r_{\alpha }(\lambda +i0).  \label{nal}
\end{equation}%
The pair $r_{\alpha }(z)$ solves the system of two coupled functional
equations
\begin{equation}
r_{a}(z)=\int_{-\infty }^{\infty }\frac{\nu _{0}(E)dE}{E+s\alpha
-z-v^{2}r_{-\alpha }(z)},\;\alpha =\pm ,  \label{rgam}
\end{equation}%
with $\nu _{0}$, defined by (\ref{non}) - (\ref{no}). The solution is unique
in the class of functions analytic for $\Im z\neq z$, and satisfying the
condition $\Im r_{a}(z)\cdot \Im z>0,\;z\neq 0$.

The $n\rightarrow \infty $ limit of the reduced microcanonical distribution (%
\ref{mer}) is then
\begin{equation}
\overline{\omega }_{\alpha \gamma }(\lambda )=\delta _{\alpha ,\gamma }%
\overline{\omega }_{\alpha }(\lambda ),\quad \overline{\omega }_{\alpha
}(\lambda )=\frac{\overline{\nu }_{\alpha }(\lambda )}{\overline{\nu }%
_{+}(\lambda )+\overline{\nu }_{-}(\lambda )}.  \label{mer2}
\end{equation}%
Correspondingly, the $n\rightarrow \infty $ limit of the reduced canonical
distribution (\ref{cer}) is
\begin{equation}
\rho _{\alpha \gamma }(\beta )=\delta _{a,\gamma }\rho _{\alpha }(\beta
),\quad \rho _{\alpha }(\beta )=\frac{\int_{-\infty }^{\infty }e^{-\beta
\lambda }\nu _{\alpha }(\lambda )d\lambda }{\sum_{\delta =\pm }\int_{-\infty
}^{\infty }e^{-\beta \lambda }\nu _{\delta }(\lambda )d\lambda }.
\label{cer2}
\end{equation}%
The absence of off-diagonal elements in $\rho (\beta )$ is due to the
special structure of the interaction (proportional to the $\sigma _{x}$
matrix). This is a rather common phenomenon in random matrix theory and many
ensembles, studied so far, have this property. If, on the other hand, one
puts an arbitrary 2 by 2 matrix instead of $\sigma _{x}$ then $\rho (\beta )$
will be non-diagonal.

In the next section we will consider models of $\mathcal{S}_{n}$ in which (%
\ref{mer2}) and (\ref{cer2}) are the same for the proper choice of $\beta $.
It will turn out that we will have not only the "equivalence of ensembles"
but also that the $\rho_\alpha(\beta)$ will be independent of the
interactions and therefore be of the Gibbs form,
\begin{equation}
\frac{e^{-\beta s\alpha }}{2\cosh \beta s}  \label{cer4}
\end{equation}%
%

\bigskip \noindent \textit{3.2. A model for the reservoir.}

\noindent\ We will consider now a model
in which $\mathcal{S}_{n}$ consists of $J$ particles or other subsystems,
each having $m$ levels (orbitals). We then have $n=m^{J}$. We assume further
that the subsystems do not interact. The (normalized) density of states $\nu
_{0}^{(n)}$ (\ref{non}) is then the $J$-fold convolution of the (normalized)
density of states $q_{m}$ of a subsystem, and we assume that in the limit $%
m\rightarrow \infty $ $q_m$ will approach some normalized density $q$.

Following the limit $m\rightarrow \infty $ we will take the limit $%
J\rightarrow \infty $. Note that this is very different from the usual
thermodynamic limit, e.g.\ for $J$ particles on a lattice of $m$ sites,
where $J$ would grow with $m$ in such a way that we would have for large $J$
that the number of levels between $E=Je$ and $E=Je+dE$ grows as $\exp
\{JS_{J}(e)\}dE$. The entropy $S_J(e)$ would then be a monotone increasing
function of $e$ which would approach, as $J\rightarrow \infty $, the
entropy/per particle, $S(e)$ with $S^{\prime }(e)=\beta $, the reciprocal
temperature of the reservoir. The same happens in spin  systems where $m$
remains finite as $J\rightarrow \infty $, in which case $S(e)$ would have a
maximum at some finite $e$, after which it would decrease.

The reason we cannot follow this procedure here is the dissonance between
random matrix theory and thermodynamic behavior already mentioned above. To
get a (normalized) density of states for the thermodynamic system would
require that $\exp \{JS_{J}(E/J)\}/m^{J}$ have a limit as $J\rightarrow
\infty $ which is not the case. 

We are therefore forced to keep $J$ fixed while taking the limit $%
m\rightarrow \infty $, yielding a continuous $q$ and $\nu _{0}$ as its $J$%
-fold convolution. This form of $\nu _{0}$ is standard in probabilities
rather than in statistical mechanics. However, they both suggest
that we consider cases where
\begin{equation}
\lim_{J\rightarrow \infty }s_{J}(e)=s(e),\quad s_{J}(e)=J^{-1}\log \nu
_{0}(Je).  \label{entr}
\end{equation}%
The function $s$, appearing above, is in fact the rate function of the large
deviation theory (see e.g. \cite{De-Ze}) and can also be seen as an analog
of the entropy of statistical mechanics, although, since $\nu _{0}$ is
normalized to unity, the analogy can not be too close. In particular, $s$
will be negative everywhere, since $\nu_0(Je) \to 0$ as $J \to \infty$,
unlike the genuine entropy in quantum statistical mechanics.

Formula (\ref{entr}) suggests the following one:
\begin{equation}
\nu _{0}(Je+\varepsilon )/\nu _{0}(Je)=\exp \{J[s_{J}(e+\varepsilon
/J)-s_{J}(e)]\}\rightarrow e^{s^{\prime }(e)\varepsilon },\;J\rightarrow
\infty .  \label{asn0}
\end{equation}%
Hence, the equality
\begin{equation}
s^{\prime }(e)=\beta ,  \label{nobet}
\end{equation}%
analogous to the well known statistical mechanics relation between the
entropy and the temperature, implies the formula
\begin{equation}
\nu _{0}(Je+\varepsilon )/\nu _{0}(Je)\rightarrow e^{\beta \varepsilon },
\label{gib0}
\end{equation}%
To obtain a positive $\beta $ we have to choose an interval on which $s(e)$ of (\ref%
{entr}) increases.

However, for the analysis of (\ref{mer2}), and (\ref{cer2}), as well as in
the study of the time dependent properties of the model, we will need the
asymptotic formula
\begin{equation}
\nu _{\alpha }(Je)/\nu _{0}(Je)\rightarrow e^{-\beta s\alpha
},\;J\rightarrow \infty ,  \label{na/n0}
\end{equation}%
where $e$ and $\beta $ are related by (\ref{nobet}). This requires certain
conditions on $q$ (in particular, the existence of certain exponential
moment) and a certain amount of technicalities. To keep our presentation
simple we consider a model case, where $q$ is the Gaussian density
\begin{equation}
q(e)=\frac{1}{(2\pi a^{2})^{1/2}}\exp \left\{ -(e-e_{0})^{2}/2a^{2}\right\} .
\label{Ga0}
\end{equation}%
In this case we have immediately
\begin{equation}
\nu _{0}(E)=\frac{1}{(2\pi Ja^{2})^{1/2}}\exp \left\{
-(E-Je_{0})^{2}/2Ja^{2}\right\} ,  \label{Gaus}
\end{equation}%
and, hence, in view of (\ref{entr}),
\begin{equation}
s(e)=-(e-e_{0})^{2}/2a^{2}.  \label{entrG}
\end{equation}%
Furthermore, we prove in Appendix 1 that if $\nu _{0}$ is given by (\ref%
{Gaus}) then the density $\nu _{\alpha }$, defined by (\ref{nal}) - (\ref%
{rgam}), verifies the asymptotic relation (\ref{na/n0}).

It is easy now to show that this model yields a reduced density matrix of ${}%
\mathcal{S}_{2}$ that is diagonal in $\alpha $ and has the same Gibbs form (%
\ref{cer4}) for both the microcanonical and the canonical cases:
\begin{equation}
\lim_{J\rightarrow \infty }\lim_{m\rightarrow \infty }\left. \frac{\overline{%
\nu }_{\alpha }^{(n)}(\lambda )}{\sum_{\gamma =\pm }\overline{\nu
}_{\gamma }^{(n)}(\lambda )}\right\vert _{\lambda
=Je}=\lim_{J\rightarrow \infty }\lim_{m\rightarrow \infty
}\frac{\int_{-\infty }^{\infty }e^{-\beta Je}\nu _{\alpha
}^{(n)}(Je)de}{\sum_{\gamma =\pm }\int_{-\infty }^{\infty
}e^{-\beta Je}\nu _{\gamma }^{(n)}(Je)de}=\frac{e^{-\beta s\alpha
}}{2\cosh \beta s},  \label{Gib1}
\end{equation}%
provided that the energy $e$ "per particle" and the temperature $\beta $ of
the reservoir are related via the usual equation (\ref{nobet}). The first
formula follows immediately from (\ref{na/n0}). To obtain the second formula
we have to take into account that (\ref{nobet}) is the saddle point
condition for the respective integrals in the $n=\infty $ version (\ref{cer2}%
) of the canonical reduced density matrix (\ref{cer}).

We see that because of the rather "non-macroscopic" (in the thermodynamic
sense) nature of the density of states the random matrix model of the
reduced equilibrium description does not exhibit, in general, properties
that are natural to expect from the statistical mechanics point of view. The
same is valid for the time evolution, as will be seen in the next sections.
However, the above ansatz for the density of states (be it given by (\ref%
{Gaus}) or something different) leads to the equivalence of the reduced
microcanonical and canonical description in the $J\rightarrow \infty $ limit
of our composite system, if we assume, in addition, a natural from the
statistical mechanics point of view relation (\ref{nobet}) between the
energy and the temperature.


\section{Time Evolution}

\label{s:evol} \noindent \textit{4.1. General formulas.}

\medskip \noindent 
Assume that the energy $E_{k}$ in (\ref{per}) belongs to a subsequence $%
\{E_{k_{n}}\}$ that converges to a given energy $E$ of the reservoir such
that $\nu _{0}(E)>0$. 
By using an extension of the techniques, presented in \cite{P1,P2}, one
obtains the following expression for the $n\rightarrow \infty $ limit of the
transfer matrix (\ref{per}):
\begin{eqnarray}
T_{\alpha \beta \gamma \delta }(E,t) &=&  \label{T} \\
&=&\frac{1}{(2\pi i)^{2}}\int_{C}\int_{C}dz_{1}dz_{2}e^{it(z_{1}-z_{2})}r_{%
\alpha \beta \gamma \delta }(E,z_{1},z_{2}),  \notag
\end{eqnarray}%
where the contour $C$ encircles the real axis,
\begin{equation}
r_{\alpha \beta \gamma \delta }(E,z_{1},z_{2})=\lim_{n\rightarrow \infty
}\left\langle \sum_{j=1}^{n}R_{\gamma k,\delta j}(z_{1})R_{\alpha j,\beta
k}(z_{2})\right\rangle ,  \label{rR}
\end{equation}%
and $R(z)=(H_{n}-z)^{-1}$ is the resolvent of the Hamiltonian (\ref{Ham}).
The "two-point" functions $r_{\alpha \beta \gamma \delta }(E,z_{1},z_{2})$
are analytic in $z_{1}$ and in $z_{2}$ outside the real axis and have the
form
\begin{eqnarray}
r_{\alpha \beta \gamma \delta }(E,z_{1},z_{2}) &=&r_{\gamma
}(E,z_{1})r_{\beta }(E,z_{2})D_{\gamma ,\beta }^{-1}(z_{1},z_{2})  \label{r4}
\\
&\times &\big(\delta _{\alpha ,\beta }\delta _{\gamma ,\delta }+v^{2}\delta
_{-\alpha ,\beta }\delta _{-\gamma ,\delta }r_{-\gamma ,-\beta }(z_{1},z_{2})%
\big)  \notag
\end{eqnarray}%
Here the "one-point" functions $r_{\alpha }(E,z),\;\alpha =\pm \;$ are
\begin{equation}
r_{a}(E,z)=\frac{1}{E+s\gamma -z-v^{2}r_{-\alpha }(z)},  \label{rkg}
\end{equation}%
$r_{\alpha }(z),\;\alpha =\pm \;$ solve system (\ref{rgam}),
\begin{equation}
D_{\gamma ,\beta }(z_{1},z_{2})=1-v^{4}r_{\gamma ,\beta
}(z_{1},z_{2})r_{-\gamma ,-\beta }(z_{1},z_{2}),  \label{D}
\end{equation}%
and
\begin{equation}
r_{\gamma ,\beta }(z_{1,}z_{2})=\int r_{\gamma }(E,z_{1})r_{\beta
}(E,z_{2})\nu _{0}(E)dE.  \label{r2}
\end{equation}%
is another two-point function.

These formulas, and (\ref{ron}) lead to the following expression for the $%
n\rightarrow \infty $ limit of the reduced density matrix
\begin{eqnarray}
\rho _{\alpha ,\delta }(E,t) &=&\frac{1}{(2\pi i)^{2}}\int_{C}dz_{2}%
\int_{C}dz_{1}e^{it(z_{1}-z_{2})}  \label{rrr} \\
&\times &\frac{r_{\alpha }(E,z_{1})r_{\delta }(E,z_{2})\rho _{\alpha ,\delta
}(0)+v^{2}r_{-\alpha }(E,z_{1})r_{-\delta }(E,z_{2})r_{\alpha ,\delta
}(z_{1},z_{2})\rho _{-\alpha ,-\delta }(0)}{1-v^{4}r_{\alpha ,\delta
}(z_{1},z_{2})r_{-\alpha ,-\delta }(z_{1},z_{2})}.  \notag
\end{eqnarray}%
Hence the limiting transition probabilities, corresponding to $\rho _{\alpha
,\alpha }(0)=\delta _{\alpha \beta }$, are
\begin{equation}
p_{\alpha ,\beta }(E,t)=\rho _{\alpha ,\alpha }(E,t).  \label{pab}
\end{equation}%


\bigskip \noindent \textit{4.2. Reduced density matrix and the transition
probabilities in the infinite time limit.}

\medskip \noindent By using the analyticity of the integrand of (\ref{rrr})
in $z_{1}$ and in $z_{2}$, we can write the following integral
representation for the diagonal entries of the reduced density matrix of the
system
\begin{eqnarray}
\rho _{\alpha ,\alpha }(E,t) &=&-\frac{1}{(2\pi i)^{2}}\int_{-\infty
}^{\infty }d\lambda _{2}\int_{-\infty }^{\infty }d\lambda _{1}e^{it\delta
\lambda }  \label{raa} \\
&\times &\frac{\delta r_{\alpha }(E)(\delta \lambda +v^{2}\delta r_{\alpha
})\rho _{\alpha ,\alpha }(0)+v^{2}\delta r_{-\alpha }(E)\delta r_{\alpha
}\rho _{-\alpha ,-\alpha }(0)}{\delta \lambda (\delta \lambda +v^{2}\delta
r_{\alpha }+v^{2}\delta r_{-\alpha })}.  \notag
\end{eqnarray}%
Here the integration path in $\lambda _{1}$ encircles $\lambda _{2}$ from
below in the clockwise direction, $\delta \lambda =\lambda _{1}-\lambda _{2}$%
,
\begin{eqnarray}
\delta r_{\alpha }(E) &=&r_{\alpha }(E,\lambda _{1}-i0)-r_{\alpha
}(E,\lambda _{2}+i0),  \label{drdr} \\
\delta r_{\alpha } &=&r_{\alpha }(\lambda _{1}-i0)-r_{\alpha }(\lambda
_{2}+i0),  \notag
\end{eqnarray}%
and $r_{\alpha }(z)$ and $r_{a}(E,z)$ are defined in (\ref{rgam}) and in (%
\ref{rkg}).

These formulas imply that the diagonal entries of the reduced density matrix
are, in the limit $t\rightarrow \infty $, given by
\begin{equation}
\rho _{\alpha ,\alpha }(E,\infty )=\int_{-\infty }^{\infty }\omega _{\alpha
}(\lambda )\Big(\sum_{\gamma =\pm }\nu _{\gamma }(E,\lambda )\rho _{\gamma
,\gamma }(0)\Big)d\lambda ,  \label{raain}
\end{equation}%
where
\begin{equation}
\omega _{\alpha }(\lambda )=\frac{\nu _{\alpha }(\lambda )}{\nu _{+}(\lambda
)+\nu _{-}(\lambda )},\;\alpha =\pm   \label{om}
\end{equation}%
is the $\varepsilon \rightarrow 0$ form of microcanonical distribution of
the small system, defined in (\ref{mer2}) with $\overline{\nu }_{\alpha
}(\lambda )$ of (\ref{nuab}), replaced by $\nu _{\alpha }(\lambda )$ of (\ref%
{nal}) - (\ref{rgam}), and
\begin{eqnarray}
\nu _{\alpha }(E,\lambda ) &=&\frac{1}{\pi }\Im r_{\alpha }(E,\lambda +i0)
\label{nalE} \\
&=&\frac{v^{2}\nu _{-\alpha }(\lambda )}{(E+s\alpha -\lambda -v^{2}\Re
r_{-\alpha }(\lambda +i0))^{2}+\pi ^{2}v^{4}\nu _{-\alpha }^{2}(\lambda )}%
,\;\alpha =\pm ,  \notag
\end{eqnarray}%
where the last equality follows from (\ref{rkg}). The functions $\nu
_{\alpha }(E,\lambda )$ are non-negative and satisfy the relations
\begin{equation}
\int_{-\infty }^{\infty }\nu _{\alpha }(E,\lambda )d\lambda =\int_{-\infty
}^{\infty }\nu _{\alpha }(E,\lambda )dE=1,\;\int_{-\infty }^{\infty }\nu
_{\alpha }(E,\lambda )\nu _{0}(E)dE=\nu _{\alpha }(\lambda ),\;\alpha =\pm .
\label{nanaE}
\end{equation}%
Relations (\ref{pab}) and (\ref{raain}) imply that the transition
probabilities are given by
\begin{equation}
p_{\alpha ,\gamma }(E,\infty )=\int_{-\infty }^{\infty }\omega _{\alpha
}(\lambda )\nu _{\gamma }(E,\lambda )d\lambda ,  \label{pin}
\end{equation}%
where $\omega _{\alpha }$ is the microcanonical equilibrium distribution (%
\ref{om}), and $\nu _{\alpha }(E,\lambda )$ is defined in (\ref{nalE}) - (%
\ref{nanaE}). This expression satisfies the normalization condition (\ref%
{psr}). %

As already remarked at the end of Section 3.1, there is no reason to expect
that $\rho (E,\infty )$ will correspond to any equilibrium state for a
general $\nu _{0}$. In fact we can show (see Section 7.1 ) that the diagonal
entries $\rho _{\alpha ,\alpha }(E,\infty )$ of (\ref{raain}) will in
general depend on $\rho (0)$. The situation is different for the model of
the reservoir, introduced in Section 3.2. Indeed, setting $E=Je(\beta )$ in (%
\ref{raain}), where $e(\beta )$ is defined by (\ref{nobet}), and (\ref{entrG}%
), we can change to the variable $\lambda =Je$, and replace asymptotically
the density $\nu _{\alpha }(Je)$ by $\nu _{0}(Je)\exp \{(e-e_{0})s\alpha
/a^{2}\}$, and the density $\nu _{\alpha }(Je(\beta ),Je)J$ by $\delta
(e+a^{2}\beta -e_{0})$ for $J\rightarrow \infty $, according to relations (%
\ref{nan0}), and (\ref{A.3}). As a result we obtain, in the infinite time
limit, the canonical Gibbs distribution form (\ref{Gib1}) of the diagonal
entries of the mean reduced density matrix of the small system with a
temperature, determined by the energy by $\mathcal{S}_{n}$. Note that the
interaction does not appear in the limit although it ensures the loss of
memory of $\rho (0)$. 


\section{Van-Hove Limit}

\label{s:vH}

\noindent In this limit the coupling constant $v$ of the system-reservoir
interaction tends to zero, the time $t$ tend to infinity while the
transition rate, given by first order perturbation in the interaction, is
kept fixed. In terms of (\ref{T}) this corresponds to letting
\begin{equation}
v\rightarrow 0,\quad t\rightarrow \infty ,\quad \tau =tv^{2}\quad \mathrm{\
fixed}  \label{cvH}
\end{equation}%
after the limit $n \to \infty$.

By using general formulas (\ref{rrr}) - (\ref{r4}), it is easy to show that
the off-diagonal entry $\rho _{+-}$ of the reduced density matrix vanishes
in the van Hove limit.

Let us consider the diagonal entries $\rho _{\alpha ,\alpha }(E,t)$ in this
limit. Changing variables to $\lambda _{2}=\lambda ,\ \delta \lambda
=v^{2}\xi $ \ in the integral representation (\ref{raa}), and using (\ref%
{drdr}), (\ref{nal}), and (\ref{nalE}), we obtain
\begin{equation}
\left. \delta r_{\alpha }(E)\right\vert _{\lambda _{1}=\lambda _{2}=\lambda
}=-2\pi i\nu _{\alpha }(E,\lambda ),\ \left. \delta r_{\alpha }\right\vert
_{\lambda _{1}=\lambda _{2}=\lambda }=-2\pi i\nu _{\alpha }(\lambda ),
\label{drvH}
\end{equation}%
where now
\begin{equation}
\nu _{\alpha }(\lambda )=\nu _{0}(\lambda -s\alpha ),\;\nu _{\alpha
}(E,\lambda )=\delta (E+s\alpha -\lambda ).  \label{nvH}
\end{equation}
The limit $\rho _{\alpha ,\alpha }^{vH}(E,\tau )$ of the diagonal entries,
is then given by
\begin{eqnarray*}
\rho _{\alpha ,\alpha }^{vH}(E,\tau ) &=&\frac{1}{2\pi i}\int_{-\infty
}^{\infty }d\lambda \int_{-\infty }^{\infty }d\xi e^{i\tau \xi } \\
&\times &\frac{\nu _{\alpha }(E,\lambda )\big[\xi -2\pi i\nu _{\alpha
}(\lambda )\big]\rho _{\alpha ,\alpha }(0)-2\pi i\nu _{-\alpha }(E,\lambda
)\nu _{\alpha }(\lambda )\rho _{-\alpha ,-\alpha }(0)}{\xi (\xi -i\Gamma
(\lambda ))},
\end{eqnarray*}%
where
\begin{equation}
\Gamma (\lambda )=2\pi (\nu _{+}(\lambda )+\nu _{-}(\lambda )),  \label{GvH}
\end{equation}%
and the integration with respect to $\xi $ is taken by encircling the origin
from below in the clockwise direction. Computing the integral with respect
to $\xi $ by residues, we get (cf Appendix 2)
\begin{eqnarray}
\rho _{\alpha ,\alpha }^{vH}(E,\tau ) &=&\int_{-\infty }^{\infty }\Big(%
\omega _{\alpha }(\lambda )\sum_{\gamma }\nu _{\gamma }(E,\lambda )\rho
_{\gamma ,\gamma }(0)  \label{rvH2} \\
&+&\frac{\nu _{\alpha }(E,\lambda )\nu _{-\alpha }(\lambda )\rho _{\alpha
,\alpha }(0)-\nu _{-\alpha }(E,\lambda )\nu _{\alpha }(\lambda )\rho
_{-\alpha ,-\alpha }(0)}{\nu _{+}(\lambda )+\nu _{-}(\lambda )}e^{-\tau
\Gamma (\lambda )}\Big )d\lambda .  \notag
\end{eqnarray}%
We see that the time independent part of the formula coincides with the
r.h.s. of (\ref{raain}), modulo the replacement (\ref{nvH}).

Now, taking into account (\ref{nvH}), we obtain finally that the reduced
density matrix in the van Hove limit is
\begin{equation}
\rho _{\alpha ,\gamma }^{vH}(E,\tau )=\delta _{\alpha ,\gamma }\rho _{\alpha
,\alpha }^{vH}(E,\tau ),  \label{rvHf0}
\end{equation}%
where
\begin{gather}
\rho _{\alpha ,\alpha }^{vH}(E,\tau )=\frac{\nu _{0}(E)}{\nu _{0}(E)+\nu
_{0}(E+2s\alpha )}\rho _{\alpha ,\alpha }(0)+\frac{\nu _{0}(E-2s\alpha )}{%
\nu _{0}(E-2s\alpha )+\nu _{0}(E)}\rho _{-\alpha ,-\alpha }(0)  \label{rvHf}
\\
+\frac{\nu _{0}(E+2s\alpha )}{\nu _{0}(E)+\nu _{0}(E+2s\alpha )}\rho
_{\alpha ,\alpha }(0)e^{-\tau \Gamma _{\alpha }(E)}-\frac{\nu
_{0}(E-2s\alpha )}{\nu _{0}(E-2s\alpha )+\nu _{0}(E)}\rho _{-\alpha ,-\alpha
}(0)e^{-\tau \Gamma _{-\alpha }(E)},  \notag
\end{gather}%
and
\begin{equation}
\Gamma _{\alpha }(E)=\Gamma (E+s\alpha )=2\pi \left( \nu _{0}(E)+\nu
_{0}(E+2s\alpha )\right) ,  \label{GvHa}
\end{equation}
with $\Gamma (\lambda )$, given by (\ref{GvH}). In a general case of an
arbitrary density of states $\nu _{0}$ of the reservoir the infinite time
part of the reduced density matrix (\ref{rvHf}) in the van Hove limit
depends on the initial density matrix $\rho (0)$ of the system, like it was
for our general formula (\ref{raain}) (see, however, Section 7.2).
Besides, the r.h.s. of formula (\ref{rvHf}) contains three "modes": the time
independent term, and the two terms with different exponentials in $t$. This
implies that the pair $(\rho _{++},\rho _{--})$ can not obey a system of two
differential equations of the first order, an expected form of the master
equation for $\rho _{\alpha ,\alpha }^{vH}(E,\tau )$

In fact, it is easy to see that if the first line of (\ref{rvHf}) is
independent of $\rho (0)$ for all values of $E$ (recall that this is a free
parameter of our model), then
\begin{equation}
\nu _{0}(E)=Ce^{\beta E},  \label{nuGi}
\end{equation}%
where $C$ and $\beta $ are constants. This form of $\nu _{0}$ leads
evidently to the Gibbs form of the equilibrium distributions (\ref{mer2}),
and to the right side of (\ref{rvHf}) having the form
\begin{equation}
\frac{e^{-\beta s\alpha }}{2\cosh \beta s}+\frac{e^{-\beta s\alpha }}{2\cosh
\beta s}\rho _{\alpha ,\alpha }(0)e^{-\tau \Gamma _{\alpha }(E)}-\frac{%
e^{\beta s\alpha }}{2\cosh \beta s}\rho _{-\alpha ,-\alpha }(0)e^{-\tau
\Gamma _{-\alpha }(E)},  \label{rvHJ}
\end{equation}%
with
\begin{equation}
\Gamma _{\alpha }(E)=4\pi \nu _{0}(E+s\alpha )\cosh \beta s.  \label{GaE}
\end{equation}%
We note, however, that while the form (\ref{nuGi}) of $\nu _{0}$ for all $E$
is incompatible with the normalization condition (\ref{no}), it can be valid
\textquotedblleft locally\textquotedblright\ in $E$ as in formula (\ref{gib0}%
). We have seen in Section 3.2 that (\ref{gib0}) implies the Gibbs form of
the equilibrium densities matrices (\ref{mer2}), and (\ref{cer2}) in the
zero coupling limit. 

Likewise, by using the asymptotic relation (\ref{gib0}) in the
time-independent term of (\ref{rvHf}), we obtain the first term of (\ref%
{rvHJ}) i.e. the Gibbs form of the time-independent term (\ref{rvHf}), after
the replacement $E$ by $Je(\beta )$ and the subsequent limit $J\rightarrow
\infty $.

The situation with the time dependent part of the reduced density matrix in
the considered regime (\ref{cvH})--(\ref{asn0}) is less satisfactory.
Indeed, according to relation (\ref{Gaus}), the density of states $\nu
_{0}(Je(\beta ))$ is exponentially small for large $J$ (except the case,
where $e(\beta )-e_{0}$ is of the order $J^{-1/2}$, that corresponds to the
"infinite temperature" $\beta =0$). Hence the decay rate (\ref{GaE}) $\Gamma$
will also be exponentially small in $J$.

A simple way
to make $\Gamma _{\alpha }(E)$ in (\ref{GvHa}) non-vanishing as $%
J\rightarrow \infty $ is to assume that the coupling constant $v$ in (\ref%
{Ham}) depends on $J$. This can be done if we write $v$ as the product of
two factors $v_{1}$ and $v_{2}$, use $v_{1}$ in the definition (\ref{cvH})
of the van Hove limit and keep $v_{2}$ in the decay rate. This yields
\begin{equation}
\Gamma _{\alpha }(E)=4\pi v_{2}^{2}(\nu _{0}(E)+\nu _{0}(E+s\alpha ))
\label{GaE1}
\end{equation}%
instead of (\ref{GvHa}) and allows us to rescale the decay rate by setting $%
v_{2}$ proportional to $(\nu _{0}(E)+\nu _{0}(E+s\alpha ))^{-1}$. Since the
parameter $E$ appears in the initial formulas (\ref{ver}) and (\ref{per})
via the condition $E_{k}=E$, we see that we have to make the interaction
term in (\ref{Ham}) dependent on the energy levels of the reservoir.
In the next section we consider a version of the model, possessing this
property.


\section{A More General Interaction Model}

\label{s:WL}

\noindent In this section we consider a model, given by the Hamiltonian (\ref%
{Ham}), in which the system - reservoir interaction $w_{n}$ is a real
symmetric Gaussian random matrix with statistically dependent entries whose
covariances are given by the relations (cf (\ref{GUE})):
\begin{equation}
\mathbf{E}\{w_{jk}\}=0,\quad \mathbf{E}\{w_{jk}w_{lm}\}=f(E_{j},E_{k})(%
\delta _{jl}\delta _{km}+\delta _{jm}\delta _{kl}),\;j,k=1,...,n,
\label{band}
\end{equation}%
where $f(E,E^{\prime })$ is a non-negative symmetric function. Assuming that
$f(E,E^{\prime })$ vanishes if $|E-E^{\prime }| > b$, for some $b$, we
obtain the so-called band matrices, whose entries are non zero inside the
"band" of the width $b$. The case $f(E,E^{\prime })\equiv 1$ corresponds to (%
\ref{GUE}). Similar models were considered in \cite{Me-Co,Pe,Bu,Lu-We}.

A somewhat tedious calculation leads to a transfer matrix $T_{\alpha \beta
\gamma \delta }(E,t)$ that is given again by formula (\ref{T}) in which now
(cf (\ref{r4}))
\begin{eqnarray}
r_{\alpha \beta \gamma \delta }(E,z_{1},z_{2}) &=&r_{\gamma
}(E,z_{1})r_{\beta }(E,z_{2})  \label{r4v} \\
&\times &\Big(1-S_{-\alpha ,-\delta }(z_{1},z_{2})S_{\alpha ,\delta
}(z_{1},z_{2})\Big)^{-1}  \notag \\
&\times &(\delta _{\alpha ,\beta }\delta _{\gamma ,\delta }+\delta _{-\alpha
,\beta }\delta _{-\gamma ,\delta }s_{-\gamma ,-\beta }(z_{1},z_{2})),  \notag
\end{eqnarray}%
where (cf (\ref{rkg}))
\begin{equation}
r_{\alpha }(E,z)=\frac{1}{E+s\alpha -z-\Delta _{-\alpha }(E,z)}.  \label{r1v}
\end{equation}%
The \textquotedblright self-energies\textquotedblright\ $\Delta _{\alpha
}(E,z),\ \alpha =\pm \ $ solve the system of two nonlinear integral
equations (cf ({\ref{rgam}))
\begin{equation}
\Delta _{\alpha }(E,z)=\int_{-\infty }^{\infty }\frac{v^{2}f(E,E^{\prime
})\nu _{0}(E^{\prime })dE^{\prime }}{E^{\prime }+s\alpha -z-\Delta _{-\alpha
}(E^{\prime },z)},\ \alpha =\pm ,  \label{Desv}
\end{equation}%
$S_{\alpha ,\delta }(z_{1},z_{2})$ is an integral operator, acting on a
function }$\phi ${\ of $E$ by the formula
\begin{equation}
\left( S_{\alpha ,\delta }(z_{1},z_{2})\phi \right) (E)=\int_{-\infty
}^{\infty }v^{2}f(E,E^{\prime })r_{\alpha }(E^{\prime },z_{1})r_{\delta
}(E^{\prime },z_{2})\phi (E^{\prime })\nu _{0}(E^{\prime })dE^{\prime },
\label{opS}
\end{equation}%
and}%
\begin{equation}
s_{\alpha ,\delta }(z_{1},z_{2})=(S_{\alpha ,\delta }(z_{1},z_{2})\mathbf{1}%
)(E).  \label{sv}
\end{equation}%
{It is easy to see that if $f(E,E^{\prime })\equiv 1$, then the above
formulas coincide with our previous formulas (\ref{r4}) - (\ref{r2}) in
which $\Delta _{\alpha }(E,z)=v^{2}r_{\alpha }(z)$. }

We show in Section 7.3 that particular asymptotic cases of the above
formulas correspond to the results of papers \cite{Me-Co,Pe}.

Consider now the van-Hove limit of this model as defined in (\ref{cvH}). One
obtains then the expressions (\ref{rvHf}) - (\ref{GvHa}) in which $\nu
_{\alpha }(\lambda )=\nu _{0}(\lambda -s\alpha )$ is replaced by
\begin{equation}
\nu _{0}(\lambda -s\alpha )f(\lambda +s\alpha ,\lambda -s\alpha ).
\label{vnvH}
\end{equation}%
Following \cite{Lu-We}, we choose the form-factor $f$ in (\ref{band}) in the
form
\begin{equation}
f(\lambda ,\mu )=\frac{w(\lambda -\mu )}{\sqrt{\nu _{0}(\lambda )\nu \nu
_{0}(\lambda )}},\;w(\lambda )=w(-\lambda ),  \label{WL}
\end{equation}%
which is in agreement with our discussion at the end of the previous
section. This leads to the following form of the diagonal entries of the
reduced density matrix (cf (\ref{rvHJ})):
\begin{equation}
\rho _{\alpha ,\alpha }(E,\tau )=\frac{e^{-\beta s\alpha }}{2\cosh \beta s}+%
\frac{e^{\beta s\alpha }\rho _{\alpha ,\alpha }(0)-e^{-\beta s\alpha }\rho
_{-\alpha ,-\alpha }(0)}{2\cosh \beta s}e^{-\tau \Gamma },  \label{vHWL}
\end{equation}%
where now (cf (\ref{GaE}))
\begin{equation}
\Gamma =4\pi w(2s)\cosh \beta s.  \label{Gagoo}
\end{equation}%
Unlike (\ref{rvHf}), obtained in Section 5 for $f\equiv 1$, the r.h.s. of (%
\ref{vHWL}) contains only two "modes", the time independent term
and the term with the exponential $e^{-\tau \Gamma }$. As a
result, $\rho _{\alpha ,\alpha }(E,\tau ), \; \alpha=\pm \;$ is
the solution of the system of two differential equations:
\begin{equation}
\frac{d}{dt}\rho _{\alpha ,\alpha }=-2\pi w(2s)\left( e^{\beta s\alpha }\rho
_{\alpha ,\alpha }-e^{-\beta s\alpha }\rho _{-\alpha ,-\alpha }\right) .
\label{meq}
\end{equation}%
Eq. (\ref{meq}) is a version of the master equation which is believed to
describe, on an appropriate time scale \cite{Ha,Ku-Co,Sp}, the approach to
equilibrium of a small system interacting weakly with a reservoir. An analog
of (\ref{meq}) was obtained in \cite{Lu-We} (see Section 6, in particular
formulas (52) - (53) there). From the probabilistic point of view (\ref{meq}%
) describes the two-state Markov process, known as the random telegraph
signal or the dichotomy process (see e.g. \cite{LGP}); see discussion after
(7.14).


\section{Some Properties of the Models}

\label{s;asym}

\noindent In this section we discuss certain, mostly asymptotic, properties
of the models, specified in Sections 2 and 6, that do not require the
special structure of the reservoir, introduced in Section 3.2.


\bigskip \noindent \textit{7.1. The dependence of the reduced density matrix
on the initial density matrix.}

\medskip \noindent We consider here the dependence of the $t\rightarrow
\infty $ limit (\ref{raain}) of the diagonal entries of the reduced density
matrix on the initial density matrix $\rho (0)$ of the small system in the
model of Section 2.

It is easy to find from (\ref{raain}) that if the $\rho _{\alpha ,\alpha
}(E,\infty )$ are independent of $\rho (0)$, then they have the form
\begin{equation}
\rho _{\alpha ,\alpha }(E,\infty )=\frac{1}{2}\int_{-\infty }^{\infty }\frac{%
\nu _{+}(E,\lambda )+\nu _{-}(E,\lambda )}{\nu _{+}(\lambda )+\nu
_{-}(\lambda )}\nu _{\alpha }(\lambda )d\lambda ,  \label{rin}
\end{equation}%
corresponding to $\rho _{\pm }(0)=1/2\;$ in (\ref{raain}), and that if the $%
\rho _{\alpha,\alpha}(E, \infty)$ are independent of $\rho (0)$ for all $E$,
then
\begin{equation}
\int_{-\infty }^{\infty }\frac{(\nu _{+}(\lambda )-\nu _{-}(\lambda ))^{2}}{%
\nu _{+}(\lambda )+\nu _{-}(\lambda )}d\lambda =0.  \label{ond}
\end{equation}%
Thus, the equality $\nu _{+}(\lambda )=\nu _{-}(\lambda )$ has to be valid
for all $\lambda $ which is impossible for the Hamiltonian (\ref{Ham}) with $%
s\neq 0$, i.e. if the small system has a nontrivial dynamics. Hence, there
will in general exist energies of the reservoir for which the $t=\infty $
reduced density matrix of the model of Section 2 depends on $\rho (0)$.


\bigskip \noindent \textit{7.2. The regime of an almost flat density of
states of the reservoir.}

\medskip \noindent There is a case where the $\rho_{\alpha,\alpha}(E, \infty)
$ are equal to 1/2 and the off diagonal entries vanish: thus corresponding
in some way to the Gibbs distribution (\ref{Gib1}) at infinite temperature $%
\beta =0$. This occurs for an \textquotedblright almost
flat\textquotedblright\ density of states $\nu _{0}$ of the reservoir. More
precisely, we assume that $\nu _{0}$ has the form (cf (\ref{Gaus}))
\begin{equation}
\nu _{0}(\lambda )=\frac{1}{a}\varphi \left( \frac{\lambda }{a}\right) ,
\label{qCa}
\end{equation}%
where $\varphi $ is a non-negative function, having unit integral, and $a$
is the biggest parameter having the dimension of energy of the problem. We
assume in addition that the parameter
\begin{equation}
\frac{\pi \varphi (0)v^{2}}{a}=\pi \nu _{0}(0)v^{2}:=A  \label{as1}
\end{equation}%
is fixed, i.e. that the coupling constant is of the order $\sqrt{a}\;$.
Hence, the regime corresponds formally to the following limiting procedure:
\begin{equation}
a\rightarrow \infty ,\;v\rightarrow \infty ,\;A=\pi \varphi (0)v^{2}/a\;\;%
\mathrm{is\;fixed}.  \label{flat}
\end{equation}%
By using the relation
\begin{equation}
\int_{-\infty }^{\infty }\frac{dx}{x-z}=\pi i\sigma (z),\;\sigma (z)=\mathrm{%
sign}(\Im z),  \label{int}
\end{equation}%
we obtain that in this regime
\begin{equation}
v^{2}r_{\alpha }(z)=Ai\sigma (z),\;r_{\alpha }(E,z)=\frac{1}{E+s\alpha
-z-Ai\sigma (z)}.  \label{vras}
\end{equation}%
Thus, according to (\ref{nal}) and (\ref{nalE}), we have (cf (\ref{nalE}))
\begin{equation}
\nu _{\alpha }(E,\lambda )=\frac{A}{\pi \lbrack (E+s\alpha -\lambda
)^{2}+A^{2}]},\ \;v^{2}\nu _{\alpha }(\lambda )=A/\pi .  \label{nnas}
\end{equation}%
Now, computing respective integrals in (\ref{raa}) (see Appendix 2), we find
that
\begin{equation}
\rho _{\alpha ,\alpha }(E,t)=\frac{1}{2}+\alpha \frac{\rho _{+,+}(0)-\rho
_{-,-}(0)}{2}e^{-\Gamma t},  \label{roas}
\end{equation}%
where
\begin{equation}
\Gamma =4A=4\pi v^{2}\nu _{0}(0).  \label{Gas}
\end{equation}%
Formula (\ref{Gas}) for the rate of the exponentially fast convergence to
the infinite time (equilibrium) limit (\ref{rTin}) is an analog of the Fermi
Golden Rule in our context.

These formulas can be obtained as a particular case of (\ref{rvHf}), and (%
\ref{GvHa}), corresponding to the replacement of all arguments in $\nu_0$ by
zeros, which is natural in view of (\ref{qCa}) and the spirit of the
approximation, according to which $a$ is the biggest parameters with the
dimension of energy, i.e., $a>>E,s\alpha$, hence $\nu_0(E \pm
2s\alpha)\simeq \nu_0(0)$. Note, however, that (\ref{rvHf}), and (\ref{GvHa}%
) correspond to the small coupling case (\ref{cvH}), while (\ref{roas}), and
(\ref{Gas}) are obtained assuming, at least formally, that $v \to \infty$
(see (\ref{flat}))
It can also be shown that in this asymptotic regime the off-diagonal entry $%
\rho _{+,-}(E,t)$ of the reduced density matrix decay exponentially fast as $%
t\rightarrow \infty $ with the rate $2(A-\sqrt{A^{2}-s^{2}})$, if $0\leq
s\leq A$, and with the rate $2A$, if $s\geq A $.

It follows then from (\ref{roas}) that in this approximation the $t \to
\infty$ limit of the reduced density matrix is
\begin{equation}
\rho (E,\infty )=\left(
\begin{array}{cc}
1/2 & 0 \\
0 & 1/2%
\end{array}
\right) ,  \label{rTin}
\end{equation}
i.e. it has the ``infinite temperature'' ($\rho (0)$-independent) Gibbs form
of (\ref{Gib1}).

Likewise, we have for the transition probabilities from (\ref{pab}), and (%
\ref{roas}):
\begin{equation}
p_{\alpha ,\delta }(E,t)=\frac{1}{2}+\frac{\alpha \delta }{2}e^{-\Gamma t},
\label{pas}
\end{equation}%
in particular,
\begin{equation}
\rho (E,\infty )=\left(
\begin{array}{cc}
1/2 & 1/2 \\
1/2 & 1/2%
\end{array}%
\right) .  \label{pTin}
\end{equation}%
Formula (\ref{roas}) is derived in Appendix 2. The derivation makes explicit
a mathematical mechanism of the exponential decay of the second term
(exponentially fast convergence to equilibrium). Namely, to compute the
double integral in (\ref{raain}) with the functions $r_{\alpha }(z)$ and $%
r_{\alpha }(E,z)$, given by (\ref{vras}), we make analytic continuation of $%
r_{\alpha }(\lambda _{1}-i0)$ to the upper half-plane and of $r_{\alpha
}(\lambda _{2}+i0)$ to the lower half-plane, which are the
\textquotedblright non-physical\textquotedblright\ sheets for these
functions. This procedure can be viewed as a simple version of analytic
continuation of the resolvent of a self-adjoint operator via the cut,
determined by its spectrum, i.e. as a "toy" case of analysis of resonances.
In the recent rigorous studies \cite{Ba-Co,Ja-Pi:97} of the return to
equilibrium this analysis was carried out by using the techniques of the
complex dilatation, applied to the Liouville operator of the composite
system.

Formula (\ref{roas}) implies that the $\rho _{\alpha ,\alpha }(E,t)$ satisfy
the system of two differential equations
\begin{equation}
\frac{d}{dt}\rho _{\alpha ,\alpha }=-\frac{\alpha \Gamma }{2}(\rho _{\alpha
,\alpha }-\rho _{-\alpha ,-\alpha }),\;\alpha =\pm .  \label{tel1}
\end{equation}%
Eq. (\ref{tel1}) describes a two-state Markov process, known as the random
telegraph signal or the dichotomy process (see e.g. \cite{LGP}). The process
assumes alternatively two values on intervals, whose lengths are independent
identically distributed random variables with the probability density $%
\Gamma e^{-\Gamma \tau }$. The equation plays the role of a master equation
for our model in the asymptotic regime (\ref{qCa}) - (\ref{as1}) of the
almost flat density of states of the reservoir. It is the $\beta =0$ version
of the equation (\ref{meq}).


\bigskip \noindent \textit{7.3. Special cases of the generalized model.}

\medskip \noindent We consider here certain asymptotic cases of the model of
Section 6, defined by (\ref{band}). Namely, we assume that the density of
states $\nu _{0}$ of the reservoir has the \textquotedblright almost
flat\textquotedblright\ form (\ref{qCa}) - (\ref{as1}), that the form-factor
$f(E,E^{\prime })$ of (\ref{band}) is
\begin{equation}
f(E,E^{\prime })=w\left( \frac{E-E^{\prime }}{b}\right) ,  \label{vb}
\end{equation}%
and that, in addition to (\ref{as1}), the inequality
\begin{equation}
a\gg b  \label{Me}
\end{equation}%
is valid. This leads to the following system of equations for the
self-energies $\Delta _{\alpha }$ of (\ref{Desv}):
\begin{equation}
\Delta _{\alpha }(E,z)=A\int_{-\infty }^{\infty }\frac{w(x/b)dx}{x+E+s\alpha
-z-\Delta _{-\alpha }(E,z)},  \label{Dex}
\end{equation}%
where $A$ is defined in (\ref{as1}). To compute the integral we take the
convenient "Lorenzian"\ form of $w$
\begin{equation}
w(x)=\frac{1}{1+x^{2}},  \label{wx}
\end{equation}%
and we obtain
\begin{equation}
\Delta _{\alpha }(E,z)=\frac{\pi Ab}{E+s\alpha -z-\Delta _{-\alpha
}(E,z)-ib\sigma (z)},  \label{DeP}
\end{equation}%
where $\sigma (z)$ is defined in (\ref{int}). Similar argument, applied to
the operator $S_{\alpha ,\delta }$ of (\ref{opS}), shows that (\ref{r4v})
has the form (\ref{r4}) in which $v^{2}r_{\alpha }$ is replaced by $\Delta
_{\alpha }$ from (\ref{DeP}).

Before analyzing the reduced density matrix in this asymptotic regime,
consider the case, where $b\gg s$, i.e. where the system dynamics is
negligible. In this case the r.h.s of (\ref{DeP}) coincides with the r.h.s.
of (\ref{vras}), thus the diagonal entries of the reduced density matrix are
given by (\ref{roas}). This result for the transition probability was
obtained in \cite{Me-Co} by using the Dyson series in powers of the
system-reservoir interaction for the evolution operator $\exp \left\{
-it\left. H_{n}\right\vert _{s=0}\right\}$ and keeping only those terms of
the respective series  that are relevant in the $n\rightarrow \infty $
limit. They considered the regime defined by (\ref{qCa}), (\ref{as1}), and
used the inequalities $a\gg b\gg s$.

On the other hand, in the regime (\ref{Me}) but with the energy $s$ of the
small system being of the same order as the range of energies $b$,
participating in the system-reservoir interaction, we obtain from (\ref{DeP}%
)
\begin{equation}
\Delta _{\alpha }(E,z)=\frac{\pi Ab}{s\alpha -ib\sigma (z)}.  \label{DeP1}
\end{equation}%
By using (\ref{DeP1}) and the above remark on the form of (\ref{r4v}), we
obtain again formula (\ref{roas}) in which $\Gamma $ of (\ref{Gas}) is
replaced by
\begin{equation}
\Gamma _{1}=\Gamma \frac{b^{2}}{s^{2}+b^{2}}.  \label{G1}
\end{equation}%
This result was obtained in \cite{Pe}, by using a more sophisticated version
of the perturbation theory, proposed in \cite{Me-Co}.


\renewcommand{\thesection}{\Alph{section}} \setcounter{section}{1} %
\setcounter{equation}{0}

\section*{Appendices}

\noindent \textit{1. Large energy tails of }$\nu _{\alpha }(\lambda)$

\medskip \noindent We will prove here the asymptotic formula (\ref{nan0}).
Recall that the density $\nu _{\alpha }$ is defined via its Stieltjes
transform
\begin{equation}
r_{\alpha }(z)=\int \frac{\nu _{\alpha }(\lambda )d\lambda }{\lambda -z}%
,\;\Im z\neq 0,  \label{ST}
\end{equation}%
by formula (\ref{nal}), and by the system (\ref{rgam}). Because of formula (%
\ref{nal}) it suffices to consider the case, where $\Im z\geq 0$. By using
the identity
\begin{equation*}
\frac{1}{\lambda -\zeta }=i\int_{0}^{\infty }e^{-i(\lambda -\zeta
)t}dt,\;\Im z>0,
\end{equation*}%
and taking into account that, according to (\ref{ST}), $\Im r_{\alpha
}(z)\cdot \Im z>0,\;\Im z\neq 0$, we can write (\ref{rgam}) for $\Im z>0$ as
\begin{equation}
r_{\alpha }(z)=i\int_{0}^{\infty }e^{-it(s\alpha -z-v^{2}r_{-\alpha
}(z))}f_{J}(t)dt,  \label{raG}
\end{equation}%
where
\begin{equation}
f_{J}(z)=\int_{-\infty }^{\infty }e^{-itE}\nu _{0}(E)dE=\exp \left\{
-ite_{0}J-\frac{Jt^{2}a^{2}}{2}\right\} .  \label{fJ}
\end{equation}%
In writing the last equality we used the form (\ref{Gaus}) of the density of
states of the reservoir. Since $\Im r_{\alpha }(E+i0)$ is non-negative for $%
\Im z\geq 0$, formulas (\ref{raG}) and (\ref{fJ}) imply the inequality
\begin{equation}
|r_{\alpha }(E+i0)|\leq \left( \pi /2Ja^{2}\right) ^{1/2}  \label{rabo}
\end{equation}%
The same formulas lead to the relation
\begin{equation}
r_{\alpha }(Je+i0)=ie^{\frac{J^{2}\xi ^{2}a^{2}}{2}}\int_{0}^{\infty }e^{-%
\frac{Ja^{2}}{2}(t-i\xi )^{2}-it(s\alpha -v^{2}r_{-\alpha }(Je+i0))}dt,
\label{raG1}
\end{equation}%
where
\begin{equation}
\xi =x/a^{2},\;x=e-e_{0}.  \label{A.1}
\end{equation}%
Replace the integral in the r.h.s. of (\ref{raG1}) by the sum of the two
integrals $I_{1}$ and $I_{2}$ over the intervals $[0,i\xi ]$ and $[i\xi
,\infty +\xi y+)$ respectively. The first integral can be written as
\begin{equation*}
I_{1}(e)=-\xi \int_{0}^{1}{e^{-\frac{J\xi ^{2}a^{2}}{2}u+\xi
g_{1}(u)(s\alpha -v^{2}r_{-\alpha }(Je+i0))}}g_{2}(u)du,
\end{equation*}%
where
\begin{equation*}
g_{1}(u)=\frac{u}{1+\sqrt{1-u}},\;g_{2}(u)=\frac{1}{2\sqrt{1-u}}.
\end{equation*}%
In view of (\ref{rabo}) the leading contribution to $I_{1}$ as $J\rightarrow
\infty $ is due to the lower integration limit $u=0$. This yields the
asymptotic formula
\begin{equation}
I_{1}(e)=-\frac{1}{Jx}-\frac{1+\xi (s\alpha -v^{2}r_{-\alpha }(e+i0))}{%
J^{2}a^{4}\xi ^{3}}+O(J^{-3}),\;J\rightarrow \infty ,  \label{I1}
\end{equation}%
where $x$ and $\xi $ are defined in (\ref{A.1}).

The integral $I_{2}$ can be transformed to the form
\begin{equation*}
I_{2}(e)=ie^{-\frac{J\xi ^{2}a^{2}}{2}}\int_{0}^{\infty }e^{-\frac{%
Ju^{2}a^{2}}{2}+(\xi -iu)(s\alpha -v^{2}r_{-\alpha }(Je+i0))}du.
\end{equation*}%
The leading contribution to $I_{2}$ is also due to the lower integration
limit $u=0$, hence we have
\begin{eqnarray*}
I_{2}(e) &=&ie^{-\frac{J\xi ^{2}a^{2}}{2}}e^{-\xi (s\alpha -v^{2}r_{-\alpha
}(Je+i0))}\int_{0}^{\infty }e^{-\frac{Ju^{2}a^{2}}{2}}du(1+O(J^{-1})) \\
&=&ie^{-\frac{J\xi ^{2}a^{2}}{2}}e^{-\xi (s\alpha -v^{2}r_{-\alpha }(Je+i0))}%
\sqrt{\frac{\pi }{2Ja^{2}}}(1+O(J^{-1}).
\end{eqnarray*}%
These formulas imply (cf (\ref{rabo})) that $r_{\alpha }(Je+i0)$ has the
asymptotic form, coinciding with the r.h.s. of (\ref{I1}), and, as a result,
we obtain that
\begin{equation}
\nu _{\alpha }(Je)=\pi ^{-1}\Im r_{\alpha }(Je+i0)=\frac{v^{2}(1+O(J^{-1}))}{%
J^{2}x^{2}}\nu _{-\alpha }(Je)+\nu _{0}(Je)e^{\xi s\alpha }(1+O(J^{-1})),
\label{A.2}
\end{equation}%
where the first and the second terms in the r.h.s. are the contributions of $%
\pi ^{-1}\Im I_{1}(Je+i0)$ and $\pi ^{-1}\Im I_{2}(Je+i0)$ respectively and $%
\nu _{0}$ is given by (\ref{Gaus}). By combining (\ref{A.2}) for $\alpha
=\pm $ and (\ref{A.1}), we obtain
\begin{equation}
\nu _{\alpha }(Je)/\nu _{0}(Je)\rightarrow \exp \left\{ (e-e_{0})s\alpha
/a^{2}\right\} ,\quad J\rightarrow \infty .  \label{nan0}
\end{equation}%
This formula and (\ref{entrG}) yield (\ref{na/n0}).

Furthermore, formulas (\ref{nalE}), (\ref{rkg}), and (\ref{A.2}) imply that
for any continuous function $\phi$ we have
\begin{equation}
\lim_{J\rightarrow \infty }\int_{-\infty }^{\infty }\phi(e)\nu _{\alpha
}(Je,Je^{\prime })Jde^{\prime }=\phi(e),  \label{A.3}
\end{equation}
i.e. that $\nu _{\alpha }(Je,Je^{\prime })J$ tends to $\delta (e-e^{\prime
}) $ as a generalized function.


\bigskip \noindent \textit{2. Derivation of formula (\ref{roas})\ }

\medskip \noindent We outline here a derivation of formula (\ref{roas}) for
the mean reduced density matrix of the small system in the regime (\ref{qCa}%
) - (\ref{as1}) of almost flat density of states. By using formulas (\ref%
{drdr}) and (\ref{vras}) for $r_{\alpha }(z)$ and $r_{\alpha }(E,z)$, we can
write (\ref{roas}) as
\begin{eqnarray*}
\rho _{\alpha ,\alpha }(E,t) &=&-\frac{1}{(2\pi i)^{2}}\int_{-\infty
}^{\infty }\frac{e^{it\lambda _{1}}}{\lambda _{1}-IA}d\lambda
_{1}\int_{-\infty }^{\infty }e^{-it\lambda _{2}}\big[(\lambda _{1}-\lambda
_{2}-2iA)\rho _{\alpha ,\alpha }(0)-2iA\rho _{-\alpha ,-\alpha }(0)\big] \\
&\times &\frac{(\lambda _{1}-\lambda _{2}-2iA)}{(\lambda _{1}-\lambda
_{2})(\lambda _{2}+IA)(\lambda _{1}-\lambda _{2}-4iA)}d\lambda _{2}.
\end{eqnarray*}%
According to (\ref{drdr}), the integrand of the internal integral in $%
\lambda _{2}$ is the limiting value of a function that is analytic in the
upper half-plane and has a cut on the real axis. We see, however, that the
function can be continued in the lower half-plane and has there simple poles
at $\lambda _{2}=\lambda _{1}-i0,\;-iA,\;\lambda _{1}-4iA$. Computing the
integral by residues, we obtain after certain transformations:
\begin{equation*}
\rho _{\alpha ,\alpha }(E,t)=\frac{1}{2\pi i}\int_{-\infty }^{\infty
}e^{it\xi }\frac{(\xi -2iA)\rho _{\alpha ,\alpha }(0)-2iA\rho _{-\alpha
,-\alpha }(0)}{\xi (\xi -4iA)}d\xi ,
\end{equation*}%
where the integration path encircles $\xi =0$ from below. The integrand can
be continued in the upper half-plane, and has there the simple poles at $\xi
=0+i0,\;4iA$. Computing the integrals by residues, we obtain (\ref{roas}).

Our argument above is a toy version of a techniques of complex dilatation,
used in \cite{Ja-Pi:97,Ba-Co} to find the resonances (poles on the second
sheet) of the Liouville operator, responsible for the exponential
convergence of the density matrix of the composite system to the Gibbs
distribution. Hence, the mathematical mechanism of the exponential
convergence in our model is similar to that of the spin-boson and related
models, studied in \cite{Ja-Pi:97,Ba-Co}.

\subsection*{Acknowledgment}

We are grateful to V. Jacksic, T. Spencer and H. Spohn for helpful
discussions. The work was supported in part by NSF Grant DMR 01-279-26,
AFOSR Grant 49620-01-1-0154. We also acknowledge the hospitality of the
Institute for Advance Study and of the H. Poincare Institute where part of
this work was done.

\end{document}